\documentclass[fleqn,usenatbib]{mnras}

\usepackage{newtxtext,newtxmath}

\usepackage[T1]{fontenc}

\DeclareRobustCommand{\VAN}[3]{#2}
\let\VANthebibliography\thebibliography
\def\thebibliography{\DeclareRobustCommand{\VAN}[3]{##3}\VANthebibliography}

\usepackage{graphicx}	
\usepackage{amsmath}	
\usepackage{bm}
\usepackage{siunitx}

\newcommand{\unitvec}{\ensuremath{\hat{\bm{n}}}}
\newcommand{\dx}[1]{\mathrm{d}{#1}\,}

\newcommand{\vecl}{{\bm{l}}}

\newcommand{\vecL}{{\bm{L}}}

\newcommand{\snr}{\ensuremath{S/N}}

\newcommand{\Tfg}{T^{\mathrm{fg}}}
\newcommand{\Tksz}{T^{\mathrm{kSZ}}}

\newcommand{\Xcmb}{X^{\mathrm{CMB}}}
\newcommand{\Ycmb}{Y^{\mathrm{CMB}}}

\newcommand{\fsky}{f_{\mathrm{sky}}}

\newcommand{\planck}{{\it Planck}}

\newcommand{\websky}{\textsc{websky}}
\newcommand{\amber}{\textsc{amber}}

\newcommand\eqn[1]{equation~\ref{#1}}

\newcommand\fig[1]{Figure~\ref{#1}}

\newcommand\sect[1]{Section~\ref{#1}}

\newcommand{\fionaedit}[1]{{\textcolor{cyan}{{#1}}}}

\title[Characterising the EoR using kSZ $\times$ CMB lensing]{Characterising the epoch of reionisation using the cross-correlation of the kSZ effect and CMB lensing}

\author[N. MacCrann et al.]{Niall MacCrann,$^{1,2}$\thanks{E-mail: nm746@cam.ac.uk}
Christopher Cain,$^{3}$
Aleksandra Kusiak,$^{2,4}$
Fiona McCarthy,$^{1,2}$
\newauthor
Alexander~Van~Engelen$^{3}$,
Darby Kramer,$^{3}$
William R. Coulton,$^{5}$
Frank J. Qu$^{6,7,8}$
\\
$^{1}$DAMTP, Centre for Mathematical Sciences, University of Cambridge, Wilberforce Road, Cambridge CB3 OWA, UK\\
$^{2}$Kavli Institute for Cosmology Cambridge, Madingley Road, Cambridge CB3 0HA, UK\\
$^{3}$School of Earth and Space exploration, Arizona State University, Tempe, AZ 85281, USA\\
$^4$Institute of Astronomy, University of Cambridge, Cambridge, CB3 0HA, UK\\
$^{5}$Department of Physics, University of Oxford, Denys Wilkinson Building, Keble Road, Oxford
OX1 3RH, UK\\
$^6$Kavli Institute for Particle Astrophysics and Cosmology, Stanford University, 452 Lomita Mall, Stanford, CA, 94305, USA\\
$^7$Department of Physics, Stanford University, 382 Via Pueblo Mall, Stanford, CA, 94305, USA\\
$^8$SLAC National Accelerator Laboratory, 2575 Sand Hill Road, Menlo Park, California 94025, USA
}

\date{Accepted XXX. Received YYY; in original form ZZZ}

\pubyear{\the\year{}}

\begin{document}
\label{firstpage}
\pagerange{\pageref{firstpage}--\pageref{lastpage}}
\maketitle

\begin{abstract}
We investigate the cross-correlation of the kinematic Sunyaev-Zeldovich (kSZ) effect with lensing of the cosmic microwave background (CMB) as a probe of the epoch of reionisation. During reionisation, bubbles of ionised electrons form around over-densities, generating temperature perturbations in the CMB via the kSZ effect, which correlate with the projected matter density field probed by CMB lensing. We demonstrate using AMBER simulations that this effect can be probed via cross-correlating the squared kSZ field with the CMB lensing potential, and that the signal is sensitive to the duration and midpoint of reionisation. We forecast that for a Simons Observatory-like experiment with $5\unit{\mu K}$-arcmin white noise, covering 40\% of the sky, the signal could be marginally detected at $S/N=2$ to $3$. Meanwhile, a futuristic experiment like CMB-HD could provide a measurement of $S/N\sim50$,  yielding informative constraints on reionisation scenarios. We investigate potential challenges in measuring the signal and explore mitigation for each: contamination from extragalactic foregrounds at low redshift, contamination of the kSZ-squared estimator by lensing, and other estimator biases.
\end{abstract}

\begin{keywords}
keyword1 -- keyword2 -- keyword3
\end{keywords}



\section{Introduction}

During the epoch of reionisation (EoR), the first stars and galaxies are sources of ionising radiation, resulting in ``bubbles'' of ionised electrons. New temperature perturbations in the observed cosmic microwave background (CMB) are generated via the kinematic Sunyaev-Zeldovich (kSZ) effect \citep{sz80}, as these ionised bubbles move along the line of sight. The kSZ is thus a potential probe of the physics of reionisation; in particular there have been attempts to use its power spectrum and trispectrum to place constraints on reionisation parameters (e.g. \citealt{reichardt21,raghunathan24,beringue25,cain25,chaubal26}), 
although separating the kSZ signal from other  small-scale foregrounds in the observed CMB temperature is very challenging. 

The kSZ temperature fluctuation field, $\Tksz(\unitvec)$, is predicted to be highly non-Gaussian, motivating the use of higher-order statistics, as proposed by \citet{smith17,ferraro18}, who define a quantity $K(\unitvec)$, the square of the (high-pass filtered) CMB temperature field.  
They propose using the angular power spectrum of $K(\unitvec)$, $C_L^{KK}$, which is then a trispectrum of the kSZ temperature fluctuation, as a probe of reionisation. Recent attempts to measure this trispectrum \citep{raghunathan24,maccrann24} have been foreground-limited but can already provide  limits on reionisation parameters. 

There is likely also information about reionisation in cross-correlations between kSZ and other probes of large-scale structure. \citet{La_Plante_2022} proposed the cross-correlation of $K$ with high redshift ($z>6$) Lyman-break galaxies observed using the upcoming Roman Space Telescope as a way to extract information on reionisation. Recently, \citet{kramer25} proposed cross-correlating $K(\unitvec)$ with the patchy optical depth field or ``patchy screening'', $\tau(\unitvec)$, also estimated from the CMB. They showed that the angular cross-power spectrum of these fields, $C_L^{K\tau}$, has sensitivity to the properties (e.g. duration) of reionisation. 

We note that cross-correlating the kSZ-squared field with a tracer of large-scale structure follows exactly the  approach  known as the ``projected-fields'' estimator \citep{dore2004,hill16,ferraro16, kusiak21,bolliet22}, which has been studied extensively for the kSZ sourced from lower redshifts (most often by using a galaxy sample as the tracer). In contrast to the reionisation epoch, when there are large spatial fluctuations in the ionisation fraction, at low redshift the Universe is effectively fully ionised, and the fluctuations in ionised electron density that source the kSZ effect are simply due to their concentration in galaxies, groups and clusters. 

In this work we propose a cross-correlation statistic that aims to probe the relation between the ionised electron and matter density fields during the EoR, by cross-correlating $K$ with the CMB lensing potential, $\phi$. CMB photons are deflected due to gravitational lensing, changing the statistical properties of the observed CMB (see \citealt{lewis06} for a review). $\phi$ is related to the matter density fluctuation field, projected along the line-of-sight with a broad redshift kernel that comfortably encompasses the redshifts of reionisation (see \fig{fig:kernels}). 
As in the case of cross-correlating with galaxies or optical depth, the kSZ temperature fluctuation, which is sourced byt the line-of-sight ionised electron momentum, $p_r(\unitvec,z) =  \delta_e(\unitvec,z) \unitvec \cdot \bm{v}(\unitvec,z)$ (where $\delta_e$ is the electron overdensity and $\bm{v}$ is the velocity field), 
will have close-to zero
cross-correlation with $\phi$. This is because  negative $p_r$ is as likely as positive $p_r$ for a given $\phi$.
By using the squared (and filtered) temperature, $K(\unitvec)$, we remove the dependence on the sign of $p_r$, and do then expect a non-zero cross-correlation with $\phi$.  

In \sect{sec:theory}, we present our notation and estimators for the kSZ-squared field, $K$, lensing potential $\phi$ and their cross-correlation $C_L^{K\phi}$. Then in \sect{sec:highzsignal} we present predictions for $C_L^{K\phi}$ from simulations, and estimate their signal-to-noise for future CMB experiments. In \sect{sec:est} we discuss various challenges in measuring the signal, for example, estimator biases and foregrounds, and propose mitigation methods. Finally, we conclude in \sect{sec.conclusion}.

\section{Notation and estimators for $C_L^{K\phi}$}\label{sec:theory}

\subsection{$K$ (kSZ-squared) }\label{sec:K}

During the EoR, the Universe is ionised within "bubbles", which have radial motions, generating temperature perturbations in the observed CMB via the kSZ effect:
\begin{equation}
    \frac{\Delta T^{\mathrm{kSZ}}}{T} (\unitvec) = -\frac{1}{c} \int \dx{\chi} g(\chi) \left[1+\delta_e(\chi\unitvec)\right] v_r(\chi\unitvec),
    \label{eq:tksz}
\end{equation}
where $\chi$ is the comoving distance, $g(\chi) = e^{-\tau}\frac{\dx{\tau}}{\dx{\chi}}$ is the visibility function, $\tau$ is the optical depth, $\delta_e(\chi\unitvec)$ is the free electron overdensity, and $v_r(\chi\unitvec) = \unitvec \cdot \bm{v}(\chi\unitvec)$ is the line-of-sight component of the peculiar velocity. 

In this work we are interested in characterising the distribution of ionised electrons 
(probed by the kSZ) with respect to the matter density field (probed by lensing). Since the kSZ perturbation depends on the radial velocity, the cross power spectrum with $\phi$
is zero on average, since positive and negative $v_r(\unitvec)$ are equally likely given some $\phi(\unitvec)$. We therefore consider here the $K(\unitvec)$ statistic defined in \citet{smith17,ferraro18}. In Fourier-space, this is defined as
\begin{equation}
    K_{\vecL} = 
    N^{KK}_{L} \int \frac{\dx{^2\bm{l}}}{(2\pi)^2} W_{\vecl} W_{\vecL-\vecl} T_{\vecl}T_{\vecL-\vecl} \label{eq:KL}
\end{equation}
with the filter function,
\begin{equation}
W_{\vecl} = \sqrt{C_l^{\mathrm{kSZ}}} / \tilde{C}_l^{TT}\label{eq:Wl}
\end{equation}
where $\tilde{C}_l^{TT}$ is the total observed CMB temperature power spectrum (i.e. including instrumental noise and foreground contributions).

The $K_\vecL$ field has ``reconstruction'' noise $N_L^{KK}$ given by (in the flat-sky approximation)
\begin{equation}
N_L^{KK} = \left[ \int \frac{\dx{^2\bm{l}}}{(2\pi)^2} W_l^2 W^2_{|\vecL-\vecl|} \tilde{C}_l^{TT}\tilde{C}_{|\vecL - \vecl|}^{TT} \right]^{-1}.
\end{equation}
Note that the power of the $\tilde{C}_l^{TT}$ in the denominator of $W_l$ means that $N_L^{KK}$ scales as $(\tilde{C}_l^{TT})^2$.

\citet{smith17,ferraro18,alvarez20} argue that the auto power spectrum of this quantity, $C_L^{KK}$ can be used to constrain reionisation parameters, and recent upper-limits have been presented by \citet{raghunathan24,maccrann24} from South Pole Telescope (SPT, \citealt{spt,sptpol,spt3g}) and the Atacama Cosmology Telescope (ACT) data respectively \citep{fowler10,thornton16,henderson16}.

We show in \fig{fig:kernels} the redshift sensitivity of the $K$-field, 
quantified as
\begin{equation}
    \frac{\dx{\bar{K}}}{\dx{z}} = \int \frac{\dx{^2l}}{(2\pi)^2} W_l^2 \frac{\dx{C_l}^{\mathrm{kSZ}}}{\dx{z}}.
\end{equation}
Here $\frac{\dx{C_l}^{\mathrm{kSZ}}}{\dx{z}}$ quantifies the relative contribution to the total kSZ power spectrum as a function of redshift and $\bar{K}$ quantifies the sky-averaged small-scale kSZ power spectrum (following the notation of \citealt{smith17}).

\subsection{CMB lensing $\phi$}\label{sec:phi}

Weak gravitational lensing remaps the CMB such that the value of the observed CMB anisotropy at $\unitvec$ is equal to the unlensed CMB  (i.e. what one would observe in the absence of lensing) at position $\unitvec+\nabla\phi(\unitvec)$ where $\phi(\unitvec)$ is the CMB lensing potential \citep{lewis06}.
We can estimate the lensing potential $\phi$ via the statistical anisotropy induced in the CMB. The standard method is to use quadratic estimators \citep{okamoto03} which reconstruct the lensing potential via its impact on the off-diagonal covariance matrix of the CMB:
\begin{equation}
    \left< X_{\vecl} Y_{\vecL-\vecl} \right> \propto \phi_{\vecL}
\end{equation}
where $X,Y \in [T,E,B]$ are the observed CMB temperature and polarisation fields.
Minimum variance quadratic estimators for $\phi$ can then be constructed via appropriately weighted and normalised
averages of such products of CMB modes. In the low-instrumental noise limit, the standard quadratic estimator is no longer optimal, motivating iterative maximum likelihood (e.g. \citealt{carron17}) or sampler-based approaches (e.g. \citealt{millea22}). 
For the forecasts presented here, we use as our baseline the standard quadratic estimator case, but consider potential gains from using more optimal lensing estimators in  \sect{sec:fid_snr}.

Since it will be useful for discussion later, we include the (flat-sky) expression for the reconstruction noise, $N^{\phi\phi,XY}_L$, for the quadratic estimator $\hat{\phi}$ from CMB maps $X,Y$, given by
\begin{equation}
N_L^{\phi\phi,XY} = \left[ \int \frac{\dx{^2\bm{l}}}{(2\pi)^2} \frac{\left[f^{XY}(\vecl, \vecL-\vecl)\right]^2}{ \tilde{C}_l^{XX}\tilde{C}_{|\vecL - \vecl|}^{YY}} \right]^{-1} \label{eq:N0phi}
\end{equation}
for the $X!=Y$ case, and twice this expression for the $X=Y$ case. Here, $f^{XY}(\vecl, \vecL-\vecl)$ are coupling functions (see e.g. \citealt{okamoto03}), and $\tilde{C}_l^{XY}$ denotes the observed power spectrum i.e. including noise and foregrounds as well as cosmological signal. 

\begin{figure}
    \centering
    \includegraphics[width=0.95\columnwidth]{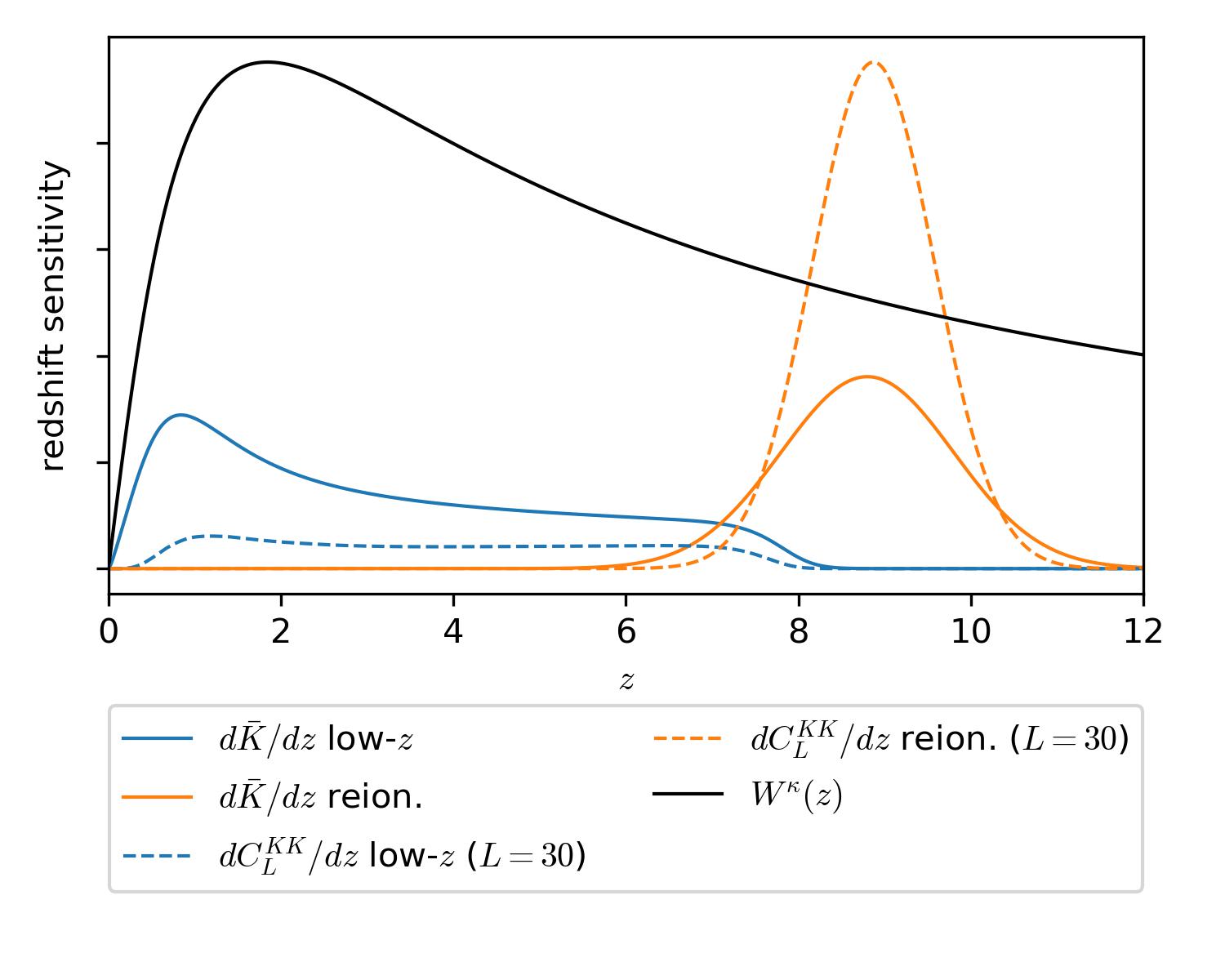}
    \caption{
    Redshift sensitivity of the $K$ and $\phi$ fields considered in this work. For $K$, we plot $\frac{\dx{\bar{K}}}{dz}$, (where $\bar{K}$ is the amplitude of the small-scale kSZ power spectrum), separated into reionisation (orange solid line) and low-$z$ (blue solid line) contributions. For comparison, we also show the redshift evolution of the contributions to $C_L^{KK}$, as discussed in \sect{sec:theory}. For $\phi$, we use the lensing kernel in \eqn{eq:lensing_kernel}. Note that the normalisation of the curves is arbitrary here, although the relative normalisation of the low-$z$ and reionisation contributions is preserved.}
    \label{fig:kernels}
\end{figure}

Unlike lensing from galaxy surveys, CMB lensing traces the matter density field all the way back to the last scattering surface, 
including therefore the redshifts, $5 \lesssim z \lesssim 20$, of interest for the EoR, as shown in \fig{fig:kernels}, where we plot the kernel\footnote{although we work with the CMB lensing potential $\phi$ throughout, this is actually the appropriate kernel for lensing convergence $\kappa$ i.e. $\kappa(\unitvec) \approx \int \dx{z} W(z) \delta_m(\unitvec, z)$, which is more directly related to the matter overdensity, and so may be more intuitive here. Given the simple relation between $\phi$ and $\kappa$ in Fourier or harmonic space, we can work with either quantity.}
\begin{equation}
    W(z) = \frac{3}{2}\Omega_m H_0^2 \frac{1+z}{H(z)}\frac{\chi(z)}{c}\frac{\chi(z_*)-\chi(z)}{\chi(z_*)}. \label{eq:lensing_kernel}
\end{equation}

\subsection{The $C_L^{K\phi}$ signal}\label{sec:clkphi}

In this work we study the cross-correlation of $K$ and $\phi$, $C_L^{K\phi}$. Given the former depends on the projected radial velocity, and the latter the projected matter over-density, we can gain intuition into the source of any signal by considering a qualitative picture where $K(\unitvec) \sim [\delta_e(\unitvec) v_r(\unitvec)]^2$ and $\phi(\unitvec) \sim \delta_m(\unitvec)$. When estimated from sufficiently small scales (high $l$) in the CMB temperature, we assume $K$ can be considered as the modulation of local, small-scale power in the ionised electron over-density, $\delta_e$, by uncorrelated linear radial velocity squared i.e. $K(\unitvec) \sim v_r^2(\unitvec)\left<\delta_e \delta_e\right>(\unitvec)$. 

For the auto-correlation of $K$, since it is sourced by large-scale correlations in $v_r^2(\unitvec)$, at least at low $L$, one can use the $\eta$-model of \citet{smith17}, which effectively drops the spatial dependence of $\left<\delta_e \delta_e\right>(\unitvec)$.
For the cross-correlation of $K(\unitvec)$ with $\phi(\unitvec)$ however, the contribution due to large-scale auto-correlations in $v_r^2$  is not present, 
and we must consider that $C_L^{K\phi}$
can arise either from correlations between $\phi(\unitvec)$ and $v_r^2(\unitvec)$, or between $\phi(\unitvec)$ and $\left< \delta_e \delta_e \right>(\unitvec)$.
While there may be some non-zero large-scale (low $L$) correlation between $\phi$ and $\left< v_r^2 \right>$, 
at sufficiently small scales (high $L$), the latter correlation will dominate, sourced by the $\left< \delta_e \delta_e \phi \right>$ bispectrum:
\begin{equation}
    C_L^{K\phi} \sim \left< v_r^2 \right> \left< \delta_e \delta_e  \phi \right> \sim \left< v_r^2 \right> \left< \delta_e \delta_e  \delta_m \right>.
\end{equation}
Given the non-Gaussianity of the ionised electron field during reionisation, we expect this bispectrum to be sensitive to the properties of reionisation. 

We direct the reader to related  arguments 
in the original projected-fields paper, \citet{dore2004}, and also recently in \citet{kramer25} in which the authors tested the analogous approximation for $C_L^{K\tau}$ i.e. that it depends on the product of $\left<v_r^2\right>$ and the $\left<\delta_e\delta_e\delta_e\right>$ bispectrum. They find it accounts for the majority of the signal on all scales, and is especially accurate at high $L$ ($L>2000)$. 
\citet{patki23} presented a more complete approach to modeling the low-redshift projected-fields signal, much of which will be applicable here, although they do not focus on the reionisation regime. 

We do not present a detailed theoretical model for $C_L^{K\phi}$  here - instead we measure the signal from simulations as well as its response to changes in reionisation parameters (see \sect{sec:amber_sims}).

Although we have noted that the $\eta$-model for $K$ of \citet{smith17} is likely not so applicable to $C_L^{K\phi}$, the redshift kernel appropriate there,  $\dx{\bar{K}}/dz$, is still probably a reasonable description of the redshift sensitivity of $K$ entering $C_L^{K\phi}$, since the relevant $\left< \delta_e \delta_e  \delta_m \right>$ bispectrum will  trace the small-scale power in $\delta_e$ as a function of $z$.

The redshift behaviours plotted in \fig{fig:kernels} indicate the likely relative size of the low and high redshift contributions to $C_L^{K\phi}$. For the auto-correlation of $K$, $C_L^{KK}$,  the more sharply-peaked $\frac{\dx{\bar{K}}}{\dx{z}}$ for the reionisation component (labelled ``$\dx{\bar{K}}/\dx{z}$ reion.''), relative to the low redshift componenet (labelled ``$\dx{\bar{K}}/\dx{z}$ low-$z$''), is advantageous, since this kernel is squared in the auto-correlation. This is apparent in the high amplitude of $\dx{C}_L^{KK}/\dx{z}$ for the reionisation component (labelled ``$\dx{C}_L^{KK}/\dx{z}$ reion.''), relative to the low redshift component (labelled ``$\dx{C}_L^{KK}/\dx{z}$ low-$z$''). 
In the cross-correlation with $\phi$ however, the broad low redshift $\frac{\dx{\bar{K}}}{\dx{z}}$ has more overlap with the lensing kernel, including its peak. As we will see in \sect{sec:lowzsignal}, this is consistent with the finding that the low-redshift contribution to $C_L^{K\phi}$ is significantly larger than the signal from reionisation. 

We assume that the covariance of the estimated signal, $\hat{C}_L^{K\phi}$, takes the form:
\begin{equation}
    \mathrm{Cov}\left[\hat{C}_L^{K\phi}, \hat{C}_{L'}^{K\phi}\right] = \frac{\delta_{LL'}}{(2L+1)f_{\mathrm{sky}}} \left[ \hat{C}_{L}^{KK} \hat{C}_L^{\phi\phi} + (\hat{C}_L^{K\phi})^2\right]\label{eq:cov_clkphi},
\end{equation}
where $\hat{C}_L^{xy}$ includes noise, $N_L^{xy}$, i.e. $\hat{C}_L^{xy} = C_L^{xy} + N_L^{xy}$ where $C_L^{xy}$ is the noiseless signal, and $\delta_{XY}$ is the Kronecker delta. Throughout, we assume the lensing potential $\phi$ is estimated only from polarisation channels.
As discussed in \sect{sec:est}, this choice removes potential estimator biases and some foreground contamination, while sacrificing little in signal-to-noise ratio ($S/N$), 
since the lensing information from future CMB experiments will be polarisation-dominated.

\section{Predicted reionisation signal from AMBER}\label{sec:highzsignal}

\subsection{\textsc{Amber} simulations}\label{sec:amber_sims}

We use the Abundance
Matching Box for the Epoch of Reionisation (\textsc{AMBER}) code \citep{trac22,chen23} to generate full-sky simulations of the kSZ signal and CMB lensing potential during reionisation.  \textsc{AMBER} uses a `semi-numerical' method, validated against radiation-hydrodynamic simulations, to generate simulations for  the reionisation history which can be directly specified via input parameters, including the midpoint and duration of reionisation, here quantified by $z_{\mathrm{mid}}$ and $\Delta z_{90}$ respectively.
For this work we generate the kSZ and lensing potential maps following the prescription in \citet{kramer25}, with our fiducial values of the input parameters following \citet{chen23}:  $z_{\mathrm{mid}} = 8$, $\Delta z_{90} = 4$, the assymetry parameter $A_z = 3$, the minimum halo mass for ionising sources, $M_{\mathrm{min}} = 10^8 M_{\odot}$, and the radiation mean-free path, $\lambda_{\mathrm{mfp}} = 3
\mathrm{Mpc}/h$.  This yields full-sky maps from which we can estimate $K$ and $\phi$. 

\subsection{Experimental setup}\label{sec:exp_setup}

As a fiducial experimental setup, we assume a white noise level of $5\mu \unit{\kelvin}$-arcmin,  a Gaussian beam with full-width-half-maximum (FWHM) 1.5 arcminute, and a fractional sky coverage $\fsky=0.4$, roughly corresponding to the expected performance for the Simons Observatory for the 93 or 145 GHz channels \citep{ade18,abitol25}. While an optimal combination of the SO frequency channels should achieve a lower effective white noise than this, foreground mitigation e.g. via additional constraints in internal linear combination (ILC) methods will increase the noise. Hence we think $5\mu\unit{K}$-arcmin is a reasonable figure.  
The total power spectra assumed to filter the maps (e.g. in \eqn{eq:Wl}), and compute the noise terms $N_L^{xy}$ in \eqn{eq:cov_clkphi} 
are 
\begin{align}
    \tilde{C}_l^{TT} &= C_l^{TT} + C_l^{\mathrm{pkSZ}} + N_l^{TT}\\ \label{eqn:cltt}
    \tilde{C}_l^{EE} &= C_l^{EE} + 2 N_l^{TT}\\
    \tilde{C}_l^{BB} &= C_l^{BB} + 2 N_l^{TT}
\end{align}
where $N_l^{TT} = \sigma^2 / b_l$, with $\sigma^2$ the  noise variance
in $(\mu$K radian$)^2$, $b_l$ the Gaussian beam function (normalised to 1 at $l=0$) and $C_l^{\mathrm{pkSZ}}$ is the reionisation contribution to the kSZ power spectrum, as measured from our baseline AMBER simulation.  In the above, the power spectra $C_l^{XY}$ without tildes are noiseless theory predictions computed with \textsc{camb} \citep{lewis11}, assuming a $\Lambda$CDM model with $\Omega_m h^2=0.120$, $\Omega_b h^2=0.022$, $H_0=67.02$, $n_s=0.9625$, $A_s=2.1509\times10^{-9}$ and $\tau=0.0657$\footnote{We note that these are slightly different to the cosmological parameters used to run the AMBER simulations, which are described in \citet{kramer25}, but this will have negligible results on the results of this paper}. We assume noise is uncorrelated between temperature and polarisation, such that $\tilde{C}_l^{TE} = C_l^{TE}$, and that $\tilde{C}_l^{TB} = \tilde{C}_l^{EB} = 0$ (due to parity symmetry conservation). We note that these total power spectra do not explicitly include sources of foreground contamination such as the thermal Sunyaev--Zel'dovich (tSZ) effect  \citep{sunyaev70,sunyaev72}, the low-redshift kSZ effect, the cosmic infrared background (CIB) \citep{puget96}, and radio sources (e.g. \citealt{zotti10,li22}). Instead we adopt a white noise level ($5\unit{\mu K}\mathrm{arcmin}$ for the SO-like case) appropriate for the increased effective noise level after performing  frequency-based removal of foregrounds e.g. a constrained internal linear combination. We will consider potential bias to the $C_L^{K\phi}$ signal due to these foregrounds in \sect{sec:est}.

We will also consider a futuristic experimental setup to investigate the potential of an experiment like CMB-HD \citep{sehgal20}; here we assume a $0.5\unit{\mu K}\mathrm{arcmin}$ white noise level and $\mathrm{FWHM}=0.3$ arcminute beam. We refer to this as the CMB-HD-like experimental setup.

For the SO-like experimental setup, we assume that the CMB lensing potential $\phi$ is estimated only from polarization, from CMB multipoles $100<l<3000$\footnote{we note that while SO data might not be sufficiently well-characterised at $l=100$, due to e.g. atmospheric noise and an uncertain transfer function, data from \planck\ could be used instead on these large scales}, and not from temperature. Meanwhile, we estimate $K$ from temperature multipoles with $3000<l<8000$.
This will remove some potential  biases on the estimator due to extragalactic foregrounds (see discussion in \sect{sec:est} and \sect{sec:fgbias}), and simplifies the estimator noise by using uncorrelated data for $K$ and $\phi$ (see discussion in \sect{sec:N0}). For the CMB-HD-like setup, we extend the CMB $l$-range to estimate $\phi$ to $100<l<8000$, but restrict to the $EB$ quadratic estimator (see also discussion in \sect{sec:est}), and we use the same $l$ range to estimate $\hat{K}$.

\subsection{Fiducial $C_L^{K\phi}$ signal and signal-to-noise}\label{sec:fid_snr}

\begin{figure}
    \centering
    \includegraphics[width=0.95\linewidth]{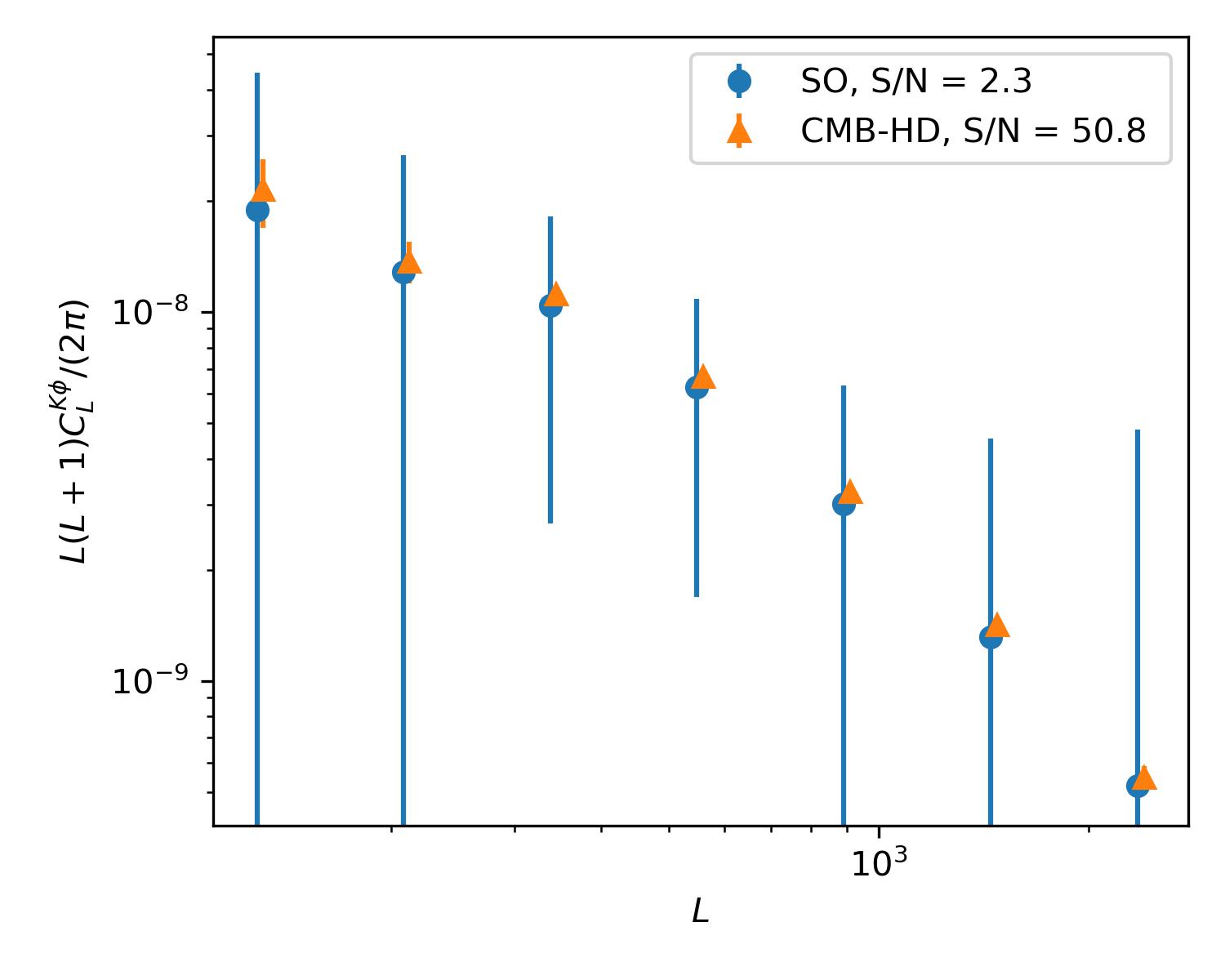}
    \caption{The cross-correlation $C_L^{K\phi}$ between $K$ (the ``kSZ-squared'' field) and the CMB lensing potential $\phi$, measured from our fiducial \textsc{AMBER} simulation. Blue and orange points show the signal under the SO-like and CMB-HD-like experimental setups respectively, with the corresponding forecast $S/N$ given in the legend (see \sect{sec:fid_snr}). Note the orange triangles are slightly offset horizontally for clarity. }
    \label{fig:fid_signal}
\end{figure}

We start by computing the $K$-field from our fiducial \textsc{AMBER} kSZ simulation, via \eqn{eq:KL}, and directly cross-correlate with the lensing potential $\phi$ map that is also output by \textsc{AMBER}. In this way, we can extract a low-noise measurement of the desired signal from the simulation (while also avoiding some complexities that arise when using a realistic $\phi$-estimator, see \sect{sec:est}). Any remaining statistical uncertainties on this estimate of the signal arise from the finite area of the \textsc{AMBER} simulation. 
We verify that after averaging over 10 independent realisations of our fiducial simulation, this uncertainty is sufficiently small for $L>100$. This signal is plotted in \fig{fig:fid_signal}. To compute expected uncertainties, we use \eqn{eq:cov_clkphi}, with the noise terms, $N_L^{xy}$, computed using \textsc{tempura}\footnote{\url{https://github.com/simonsobs/tempura}}. 

\begin{figure*}
    \centering
    \includegraphics[width=0.95\textwidth]{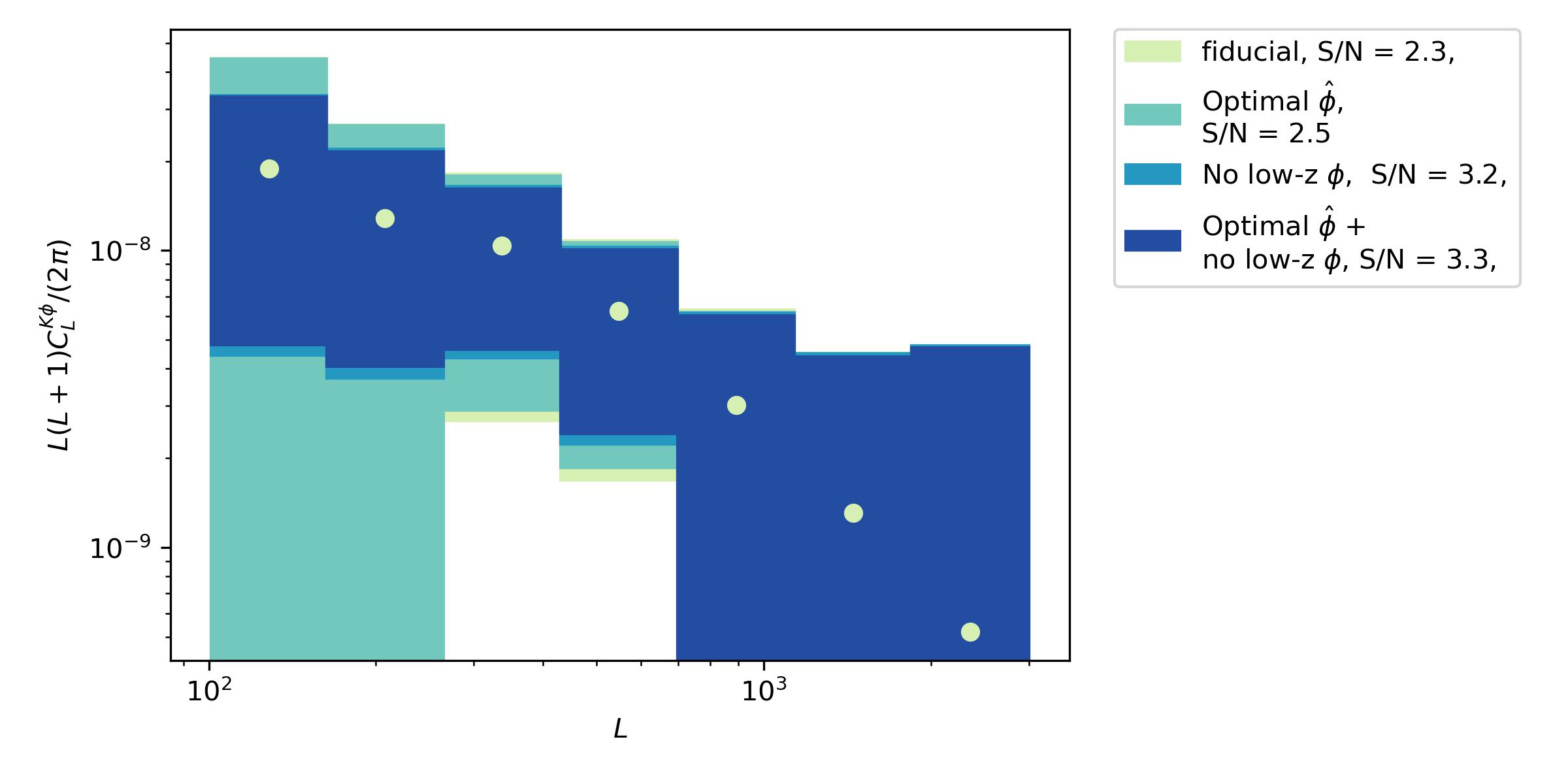}
    \caption{
    Light green points show the cross correlation between $K$ (``kSZ-squared") and CMB lensing potential $\phi$, measured from the fiducial AMBER simulation (described in \sect{sec:highzsignal}\fionaedit{)}. Fiducial uncertainties, in light green,
    assume a Simons Observatory-like experimental setup with 5$\mu\unit{K}$-arcmin white noise, a 1.5' beam, and $\fsky=0.4$. Three other sets of uncertainties show more optimistic scenarios (described in detail in \sect{sec:fid_snr}), with darker and bluer tones indicating greater constraining power: ``Optimal $\hat{\phi}$" assumes a lower lensing reconstruction noise, appropriate for an optimal estimator. ``No low-z $\phi$'' assumes no cosmic variance from low-redshift lensing. ``Optimal $\hat{\phi}$ + no low-z $\phi$'' combines both of these aforementioned scenarios.}
    \label{fig:amber_theory}
\end{figure*}

\begin{figure}
    \centering
    \includegraphics[width=0.95\columnwidth]{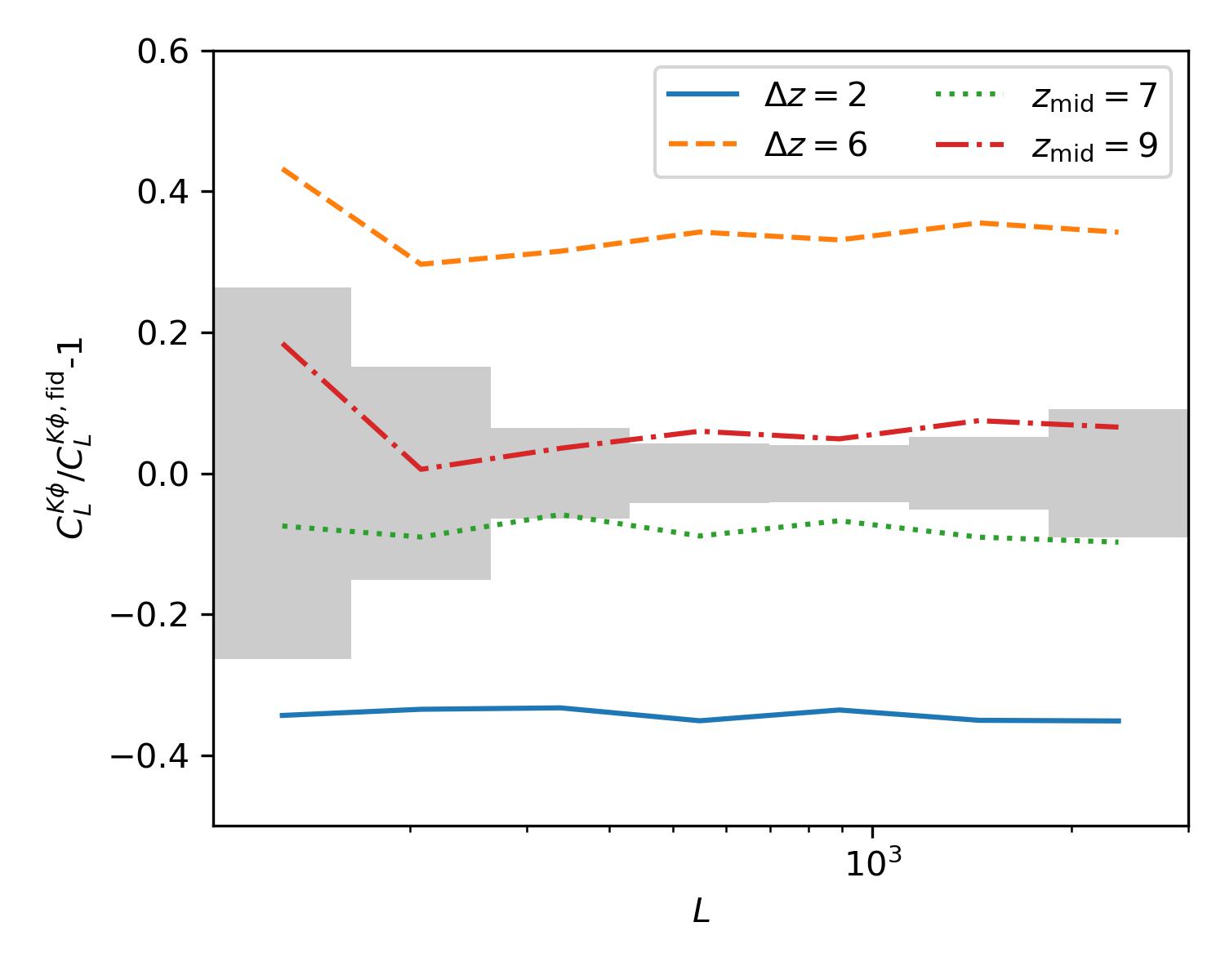}
    \caption{Sensitivity of $C_L^{K\phi}$ to variations in the reionisation 
    parameters used for the AMBER simulations. Plotted is the fractional change in $C_L^{K\phi}$ with respect to our fiducial simulation with $\Delta z = 4$, $z_{\mathrm{mid}} = 8$. We see a fractional difference of $\sim 30\%$ when the length of reionisation, $\Delta z$, is increased to 6 and a smaller increase when the midpoint of reionisation, $z_{\mathrm{mid}}$, is set to 9. 
    Grey boxes indicate uncertainties for our CMB-HD-like experimental setup. 
    }
    \label{fig:amber_variations}
\end{figure}

With the fiducial SO-like uncertainties, the signal is barely detectable, with signal-to-noise ratio of 2.3, although as in the case of $C_L^{KK}$, an upper-limit could still provide useful constraints on reionisation models. For a futuristic CMB-HD-like experiment we can expect much higher $S/N=51$. 

To get a sense of the sources of uncertainty on the signal for the SO-like experimental setup, we present in \fig{fig:amber_theory} uncertainties and $\snr$ estimates for several other cases. 

Firstly, in our baseline approach we have assumed reconstruction noise for $\hat{K}$ and $\hat{\phi}$ appropriate for simple quadratic estimators. Lower lensing reconstruction noise should be achievable using more optimal estimators e.g. iterative maximum likelihood approaches \citep{carron17} or sampling-based approaches \citep{millea22}. 
The  lensing reconstruction noise for these optimal estimators is well-approximated \citep{carron17,hotinli22} by replacing in \eqn{eq:N0phi} the observed CMB B-mode\footnote{Why only the B-mode power? Since the B-mode power is effectively  generated by lensing only, it is very effectively reduced in amplitude by delensing, leading to a significant reduction in lensing reconstruction noise. This is not the case for the TT, TE and EE cases, for which the non-zero unlensed power spectrum limits the impact of delensing on the lensing reconstruction noise.} power spectrum, $\tilde{C}_l^{BB}$,
with the  post-delensing version, $\tilde{C}_l^{BB,\mathrm{delensed}} = A_{\mathrm{lens}}C_l^{BB} + N_l^{BB}$. Here $A_{\mathrm{lens}}$ is the delensing efficiency, expected to be around 0.3 for SO \citep{namikawa22} and 0.1 for CMB-HD \citep{macinnis24}. For the SO-like case, there is only a small increase in constraining power to $S/N=2.5$. For the CMB-HD case, the gain is more significant, given the dominance of the EB lensing estimator and greater delensing efficiency, from $S/N=51$ to $S/N=62$.

In theory, the reconstruction noise for $\hat{K}$ could also be reduced via delensing, however, \citet{hotinli22,macinnis24} show that in the high-$l$ regime relevant for the $K$ estimator, delensing has rather limited impact on the temperature power spectrum, so we do not consider this case further here. 

Next we imagine that the cosmic variance contributions to the reconstructed $\phi$ can be reduced by removing the low redshift lensing using tracers of the large-scale structure e.g. galaxy surveys), as suggested in \citet{mccarthy21,qu23,baleatolizancos23}. In this case we replace $C_L^{\phi\phi}$ in \eqn{eq:cov_clkphi} with the contribution from reionisation redshifts only, i.e. the $C_L^{\phi\phi}$ measured from the AMBER simulation (labelled ``No low-$z$ $\phi$'' in \fig{fig:amber_theory}). Again, there is a modest increase in $S/N$ to 3.2. Combining these two scenarios leads to a $S/N$ of 3.3. 

Ultimately, this is a small signal - unlike the kSZ trispectrum, $C_L^{KK}$, the signal does not pick up large-scale velocity-squared correlations.
Nonetheless, any observations of the large-scale structure statistics of the EoR (e.g. from 21 cm) are extremely challenging, so alternative observables such as this one are worth considering. It will have somewhat different systematics (e.g. dependence on foregrounds) to $C_L^{KK}$, and as for that statistic, in the absence of detection, upper-limits on the signal may still provide useful limits on reionisation physics. 

\subsection{Variations in reionisation model}

\fig{fig:amber_variations} show fractional differences in the $C_L^{K\phi}$ signal, with respect to our fiducial simulation, for several variations in the AMBER simulation parameters. Those labelled $\Delta z_{90}=2$ and $\Delta z_{90} =6$ have a reduced/increased width of reionisation in redshift, while those labelled $z_{\mathrm{mid}}=7$ and $z_{\mathrm{mid}}=9$ reduce/increase the mid-point of reionisation. 

In general, we see a similar trend as identified in \citet{maccrann24}, which examined $C_L^{KK}$ for AMBER variations: when more reionisation occurs at higher redshift (due to increased $\Delta z$ or $z_{\mathrm{mid}}$), the signal increases, since ionised bubbles of a given size have higher mean density. For our fiducial experimental setup, the $C_L^{K\phi}$ signal is too weak to discriminate between these reionisation scenarios, e.g., in a Universe with true $\Delta z =4$, it would only rule out $\Delta z_{90} = 2$ at $1\sigma$, assuming all other parameters are held fixed. 

For the ambitious CMB-HD-like experimental setup,  $C_L^{K\phi}$ has much greater discriminating power - ruling out $\Delta z_{90} =2 (6)$ at $13.6 (11.9) \sigma$. We leave for future work whether such a high $S/N$ measurement of $C_L^{K\phi}$ may be able to constrain further properties of the reionisation, such as the asymmetry of the reionisation history, or even properties of the ionising sources such as their minimum halo mass.

\section{Estimator biases and mitigation}\label{sec:est}

Both the kSZ-squared, $K$, and CMB lensing potential, $\phi$, are estimated from the induced statistical anisotropy in the observed CMB, here via quadratic estimators. The cross-correlation $C_L^{K\phi}$ is then a tripsectrum of observed CMB fields. We denote a CMB field by one of $X,Y,U,V \in [T,E,B]$, such that 
\begin{align}
    \hat{C}_L^{K\phi} &\sim \left< \hat{K}_{\vecL}^* \hat{\phi}_\vecL \right>\\
    &\sim \left< \hat{K}_\vecL[X_{\vecl_1} Y_{\vecl_2}] \hat{\phi}_\vecL[U_{\vecl_3} V_{\vecl_4}] \right>\\
    &\sim \left< \hat{K}_\vecL[T_{\vecl_1} T_{\vecl_2}] \hat{\phi}_\vecl[U_{\vecl_3} V_{\vecl_4}] \right>
\end{align}
where in the third line we have substituted $X=Y=T$ (since only CMB temperature and not polarization is sensitive to the kSZ effect). We use a notation $\hat{Q}_\vecL[X_{\vecl_1}, Y_{\vecl_2}]$ to denote a quadratic estimator (e.g. for $\hat{K}_\vecL$ or $\hat{\phi}_\vecL$) operating on two CMB modes $X_{\vecl_1}$ and $Y_{\vecl_2}$. 

Various different trispectra, beyond the $C_L^{K\phi}$ signal we are interested in, arise:
\begin{enumerate}
    \item There is a bias from the disconnected trispectrum of (Gaussian) primary CMB anisotropies. Noise (either Gaussian or non-Gaussian), e.g. from the atmosphere or the instrument, will also contribute a measured trispectrum. We refer to these contributions as the $N^0$ bias and discuss them in \sect{sec:N0}. 
    \item There is  a low-redshift component of the kSZ that will correlate with the CMB lensing potential which we discuss  in \sect{sec:lowzsignal}.
    \item The $K$ estimator may also pick up statistical anisotropy from other low-redshift extragalactic foregrounds such as  the tSZ and the CIB, all of which have spatial correlations with the lensing potential $\phi$. This will generate a contamination to $C_L^{K\phi}$ that depends on a bispectrum $\left<\hat{K}[\Tfg,\Tfg] \phi\right>$ (cf. the ``primary bispectrum" bias in CMB lensing, see e.g. \citealt{vanengelen14,osborne14}), where $\Tfg$ denotes the contribution to the observed CMB temperature from foregrounds. We discuss this in \sect{sec:fgbias}.  
    \item There are also potential contamination terms if both the $\hat{K}$ and $\hat{\phi}$ estimates are affected by foregrounds - either via a foreground trispectrum of the form $\left<\hat{K}[\Tfg,\Tfg] \hat{\phi}[\Tfg,\Tfg]\right>$ or a trispectrum of the form $\left <\hat{K}[\Tfg,\Xcmb]\hat{\phi}[\Tfg,\Ycmb]\right>$, analogous to the secondary bispectrum bias in CMB lensing \citep{vanengelen14,osborne14}. These are likely to be small if $\phi$ is estimated from polarisation. 
    \item The $\hat{\phi}$ estimator is potentially biased by kSZ, leading to a trispectrum of the form $\left<\hat{K}[\Tksz,\Tksz] \hat{\phi}[\Tksz,\Tksz]\right>$. Using a polarization-only $\hat{\phi}$ estimator will remove this bias.
    \item Finally, the $\hat{K}$ estimator picks up statistical anistropy generated by lensing as well as that from the kSZ effect. This will thus generate a contamination proportional to $C_L^{\phi\phi}$. We call this the lensing bias and discuss it in \sect{sec:lensingbias}.
\end{enumerate}

\subsection{Gaussian and noise bias, $N^0$}\label{sec:N0}

 The estimator for $C_L^{K\phi}$  picks up what we refer to (following the CMB lensing convention) as an ``$N^0$ bias'' (pronounced ``N-zero''), from 
 \begin{enumerate}
 \item the disconnected trispectrum terms of the form
 \begin{equation}
     \left <\hat{K}[T_{\vecl}^{CMB},T_{l'}^{CMB}]\hat{\phi}[X_{\vecl''}^{CMB},Y_{\vecl'''}^{CMB}]\right> \sim \delta_{\vecl \vecl'} \delta_{\vecl''\vecl'''} 
     C_l^{TT,\mathrm{CMB}} C_{l''}^{UV,\mathrm{CMB}}
 \end{equation} 
 i.e. the bias that would arise for statistically isotropic CMB (without lensing or kSZ).  
 \item the noise trispectrum $\left< \hat{K}[n^{T}_{l_1}, n^{T}_{l_2}] \hat{\phi}[n^{U}_{l_3} n^{V}_{l_4}] \right>$ (where $n^X$ indicates the instrumental noise realisation  for observed channel $X$),  which can have non-zero connected and disconnected terms (the latter in the case of statistically anisotropic noise). 
 \end{enumerate}
 This $N^0$ bias can be estimated from simulations using the realization-dependent $N^0$ (RD$N^0$) \citep{dvorkin09, hanson11, namikawa13}, and subtracted from the measured signal. This is also the approach taken for $C_L^{KK}$ in \citet{maccrann24}.

We note that for the $C_L^{K\phi}$ cross-correlation, one can avoid this noise term, $N_L^{0,K\phi}$, by using independent data for the $K$ and $\phi$ estimates. There are a couple of different strategies one could use to achieve this. 

Since the primary CMB is (very close to) Gaussian, one can remove the contribution of Gaussian CMB to the $N_L^{0,K\phi}$ by using a different range of CMB $l$ for the $\hat{K}$ and $\hat{\phi}$ estimates\fionaedit{,} e.g. 
\begin{align}
    f^K(l_1,l_2) = &0 \text{ for $l_1<l_{\mathrm{cut}}$ or $l_2<l_{\mathrm{cut}}$ and}\\
    f^\phi(l_1,l_2) = &0 \text{ for $l_1>l_{\mathrm{cut}}$ or $l_2>l_{cut}$}.
\end{align}
Using  this approach with $l_{\mathrm{cut}} \approx 3000$ is a natural option for $C_L^{K\phi}$, since $K$ is mostly informed by CMB temperature modes $l>3000$ where the kSZ starts to dominate over the primary CMB, and $\phi$ is generally estimated from CMB modes $l<3000$ where the (lensed) primary CMB dominates. 
However, other sources of noise, e.g. instrumental and atmospheric noise, may be non-Gaussian (i.e. have non-zero correlations between different multipoles). Using a cross-correlation based estimator \citep{madhavacheril20}, in which independent splits of the data are used in each trispectrum leg, would remove these non-Gaussian  noise terms. 

Alternatively, if one can assume that the temperature and polarization noise are independent, this $N_L^{0,K\phi}$ can then be avoided by using only polarization information for the $\phi$ estimator i.e. $U,V \in [E,B]$. In fact if one uses only the $EB$ estimator for $\hat{\phi}$, one can use the full-range of $l$ for $f^\phi$, since the Gaussian CMB contribution will depend on $C_l^{TB}$ or $C_l^{EB}$ which are both zero. As instrumental noise decreases in upcoming CMB datasets, polarization, and in particular the $EB$ estimator, will start to dominate the CMB lensing reconstruction $S/N$, so the loss of $S/N$ in $C_L^{K\phi}$ from this strategy will not be too large (see Sections~\ref{sec:highzsignal} \& \ref{sec:lowzsignal}). 

As discussed in \sect{sec:exp_setup}, we decide on using polarisation data only for the $\hat{\phi}$ estimation here. While this results in a small reduction in $S/N$\footnote{Even if the $N^{0, K\phi}$ is ignored, and the full range of scales $l$ in CMB temperature is used in the $\phi$ estimator, the $S/N$ for the SO-like setup only increases to 3.2} compared to using temperature data as well, there is an additional motivation beyond simplifying the estimator by removing the $N^0$ bias: we also remove the bias due to the foreground trispectrum $\left<\hat{K}\left[\Tfg\Tfg\right]\hat{\phi}\left[\Tfg\Tfg\right]\right>$, which we believe is well-worth this small reduction in $S/N$. 

\subsection{Low-redshift signal}\label{sec:lowzsignal}

As well as during reionisation, the kSZ is sourced at lower redshifts, via the radial velocities of over-dense regions of ionised electrons in galaxies, groups and clusters. As in the ``projected-fields'' kSZ estimator, this kSZ signal can also be probed using a temperature-squared estimator, and then cross-correlated with large-scale structure tracers. Since the tracer considered here, CMB lensing, probes the structure at low redshifts as well, then without mitigation, this low redshift signal will be a significant contaminant to the reionisation signal. 

\begin{figure}
    \centering
    \includegraphics[width=0.95\columnwidth]{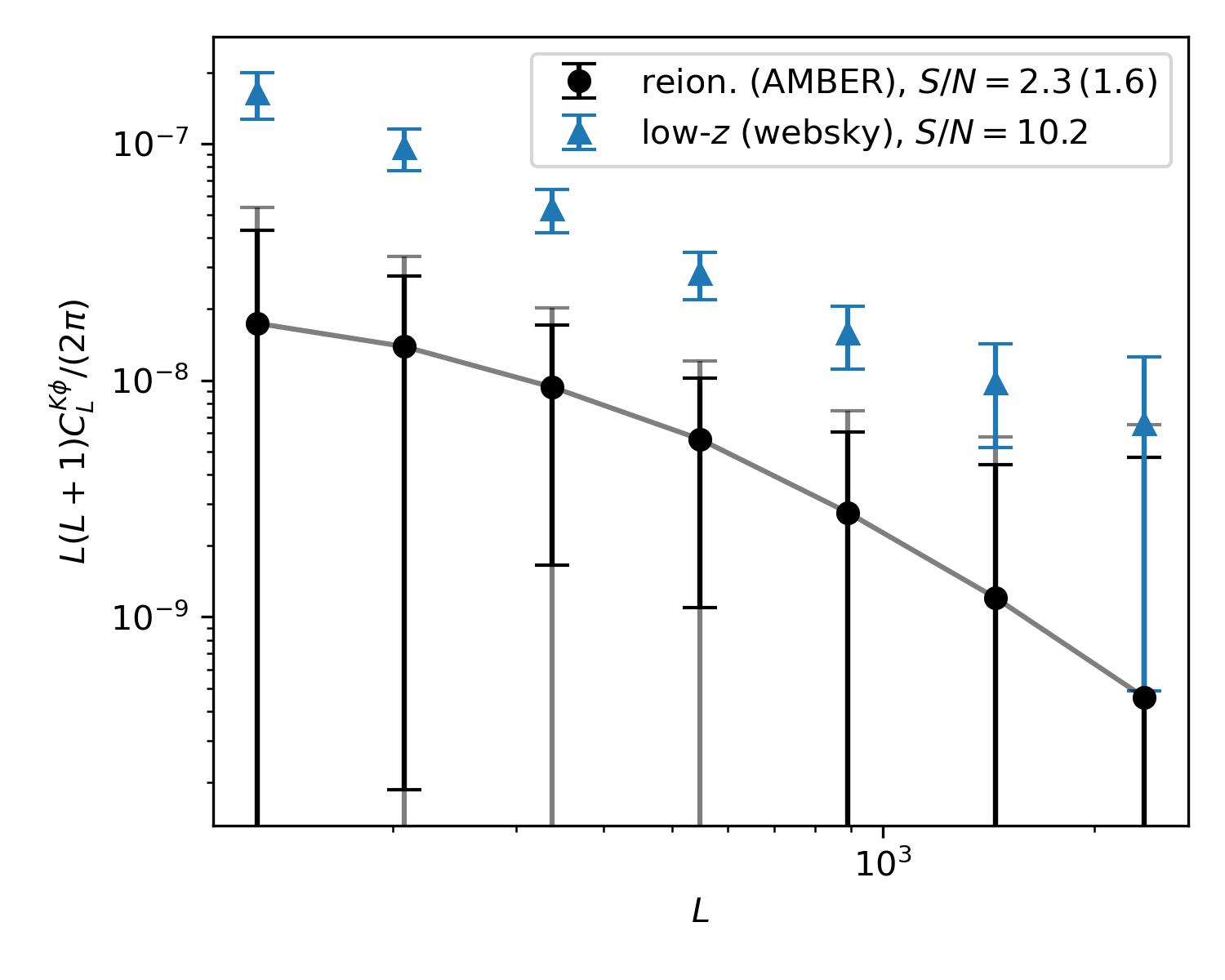}
    \caption{Blue traingles show the low redshift contamination to $C_L^{K\phi}$, as estimated from the \websky\ simulation. Without mitigation, it exceeds the reionisation signal from the fiducial \amber\ simulation (black circles) by a factor between 5 and 10. The $S/N$ for the two signals is indicated in the legend. For the \textsc{AMBER} simulation we show in parentheses the $S/N$ when including low-redshift kSZ power in the covariance calculation, which results in the grey errorbars (see \sect{sec:lowzsignal} for details). }
    \label{fig:lowz_ksz}
\end{figure}

\fig{fig:lowz_ksz} shows that the low redshift signal, predicted from the \websky\ simulations \citep{websky}\fionaedit{,} is larger than the reionisation signal at all $L$: roughly by a factor of $10\,(5)$ at $L=100 \,(1000)$. 
In the absence of other contaminating foregrounds, it would be detected at high $S/N$ in our fiducial experimental setup, and may be an interesting astrophysical signal, likely sensitive to the details of galaxy formation and feedback.   
\citet{bolliet22} also forecast the cross-correlation of low-$z$ kSZ$^2$ with CMB lensing, predicting $S/N=16$ for Simons Observatory. We note various differences between their predictions and ours: firstly, they include the lensing bias to the  kSZ$^2$ estimator, which generates additional signal when cross-correlated with the lensing. We do consider this lensing bias in \sect{sec:lensingbias} and suggest using a ``lensing-hardened'' estimator to reduce it. In addition, we use only polarization for the lensing estimator, whereas \citet{bolliet22} also include temperature (without accounting for the resulting noise correlations between the kSZ$^2$ and lensing estimators). Finally, they estimate the signal via a halo model approach whereas we use the \websky\ simulation. 

For the purposes of recovering the reionisation signal, some mitigation of the low redshift signal would clearly be necessary. Low-redshift tracers of the large-scale structure 
such as galaxies and the CIB could be used to subtract the low redshift component from the $\phi$ map, as suggested in \citet{mccarthy21,qu23} and \citet{baleatolizancos23}. \citet{qu23} forecast that the (total) CMB lensing power spectrum could be reduced by a factor of 6 at $L=100$ using LSST galaxies at $z<5$. 

This factor would roughly correspond to the reduction in the bias to our signal from low-$z$ kSZ, as well as the other extra-galactic foregrounds investigated in \sect{sec:fgbias}. The actual factor depends on the redshift kernels of the various foregrounds, and how effective the removal of the $\phi$ signal is as a function of redshift -- we do not attempt a more precise estimate in this work, but note with some optimism that a factor of 6 reduction would bring the low redshift kSZ contamination down to an order unity fractional bias, at which point theoretical modeling of the remaining bias may be sufficiently accurate to make useful inference on reionisation parameters. 

Similarly, one could also attempt to remove the low-$z$ contribution to $C_L^{K\phi}$ by removing the low-$z$ contribution to the kSZ-squared map. \citet{foreman23} present a methodology for ``de-kSZing'' i.e. using galaxy catalogs to construct a template for the late-time kSZ that can be subtracted from the CMB temperature map. They find that a $\sim10-20\%$ reduction in the kSZ power spectrum can be achieved using DESI, with the potential for a $40\%$ reduction when using LSST and other upcoming datasets. It is likely then, that enough of the low-$z$ signal would remain to require detailed joint modeling of the low-$z$ signal with the reionisation signal. 

As well as contaminating the signal, the presence of low-$z$ kSZ will increase the variance in the temperature map used for the $\hat{K}$ estimate. 
This will increase the covariance of the $C_L^{K\phi}$ signal, as can be seen straightforwardly from \eqn{eq:cov_clkphi}, where the presence of low-redshift kSZ will increase $\hat{C}_L^{KK}$ and the total $\hat{C}_L^{K\phi}$. If we assume this additional power is not accounted for in our assumed effective noise level of $5\unit{\mu K}\mathrm{arcmin}$, and add the low-redshift kSZ power as an additional source of noise, the grey errorbars in \fig{fig:lowz_ksz} result, corresponding to a decrease in the $S/N$ of our fiducial signal from 2.3 to 1.6. 

\subsection{Other foreground biases}\label{sec:fgbias}

The $\hat{K}$ statistic will also pick up contributions from other extragalactic foregrounds, such as the thermal SZ, the CIB, and radio sources (as explored in e.g. \citealt{kusiak21,kusiak23} for the low-redshift kSZ$^2$ statistic).
Migitation at the map-level via e.g. match-filter finding and subtraction of sources and clusters, or frequency-based cleaning approaches, will reduce the presence of these foregrounds in the data, but  residuals will remain. 
If the low-redshift lensing cannot be fully removed from the $\hat{\phi}$ used in $C_L^{K\phi}$ (see \sect{sec:lowzsignal} for discussion of this), then a bias to the signal will be present due to the correlation of these residuals with the lensing potential. 

Extragalactic foregrounds are spatially correlated with $\phi$, hence a term of the form $\left< K[T^\mathrm{fg}, T^\mathrm{fg}] \phi \right>$ (where $T^\mathrm{fg}$ denotes the foreground contribution to the temperature measurement), analogous to the primary bispectrum bias in CMB lensing, will be present. If the $\hat{\phi}$ estimate is also contaminated by extragalactic foregrounds, then additional contamination terms can be present, such as $\left< K[T^{\mathrm{fg}}T^{\mathrm{fg}}] \phi[T^{\mathrm{fg}}T^{\mathrm{fg}}]\right>$ (cf. the trispectrum term in CMB lensing) and $\left<K[T^{\mathrm{fg}}T^{\mathrm{cmb}}] \phi[T^{\mathrm{fg}}X^{\mathrm{cmb}}]\right>$ (cf. the secondary bispectrum term in CMB lensing). Here, we assume the foreground contamination of $\hat{\phi}$ can be neglected, mitigated by the various methods available such as bias-hardening or multi-frequency cleaning, or by using using only polarization data for the $\hat{\phi}$ estimator (in which extragalactic foregrounds are presumed to be negligible). Then we only need consider the $\left< K[T^\mathrm{fg}, T^\mathrm{fg}] \phi \right>$ term.
\begin{figure}
    \centering
    \includegraphics[width=0.99\columnwidth]{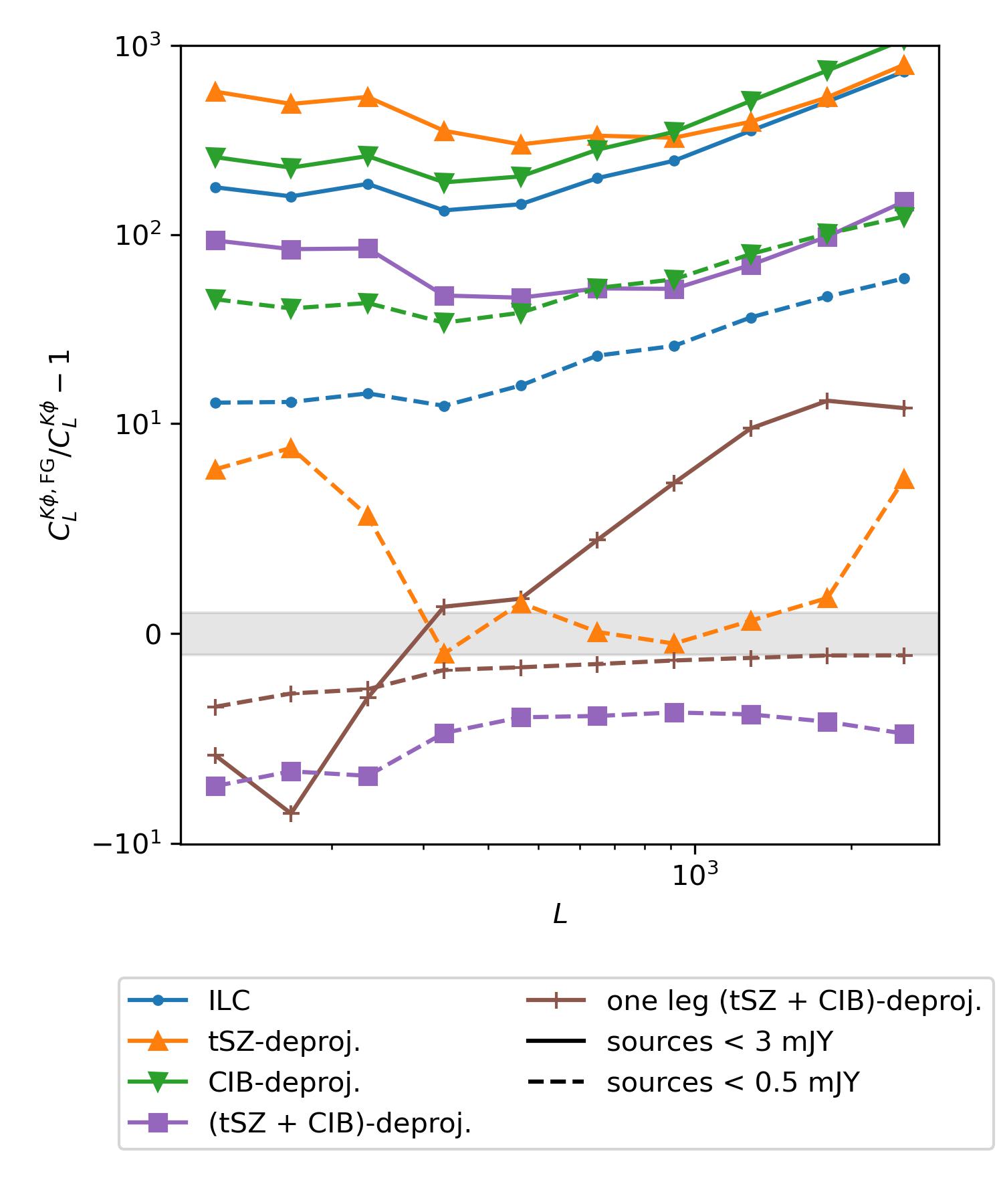}
    \caption{Fractional biases to $C_L^{K\phi}$ due to the extragalactic foregrounds tSZ, CIB and radio point sources (i.e. not including low-redshift kSZ), estimated from the \websky\ simulations. The y-axis is scale is linear in the range $[-10,10]$ 
    and logarithmic otherwise. As a guide the grey shaded region indicates $[-1,1]$. We assume an Simons Observatory-like experimental setup, see \sect{sec:fgbias} for details. For solid lines we assume point sources with flux $>3 \unit{mJY}$ at 145 GHz can be removed; dashed lines assume a more optimistic threshold of $0.5 \unit{mJY}$. For a simple harmonic ILC of the four  frequencies, the foreground bias is large, greater than two orders of magnitude at low and high $L$. While de-projecting (frequency) spectral templates for either the tSZ or CIB alone is not helpful, deprojecting both, from one or both ``legs" (i.e. temperature maps) entering the quadratic estimator $\hat{K}$, reduces the contamination significantly. 
    Since these foregrounds are almost completely sourced from redshifts lower than the epoch of reionisation, these biases would be significantly reduced by ``cleaning" low-redshift lensing using e.g. the CIB or galaxy surveys - see discussion in \sect{sec:lowzsignal}. 
    }
    \label{fig:fgbias}
\end{figure}

\fig{fig:fgbias} shows the fractional biases to $C_L^{K\phi}$ predicted from the \websky\ simulations. $K$ is estimated from the \websky\ maps after performing a harmonic-space internal linear combination (ILC, \citealt{bennett92}) for an Advanced SO-like setup: we assume white noise levels of $\left[3.8, 4.1, 10., 25 \right]\unit{\mu K arcmin}$ and beam FWHM $\left[2.2, 1.4, 1., 0.9 \right]\unit{arcmin}$ for frequencies $\left[93, 145, 225, 278\right]\unit{GHz}$ respectively. We include CIB, tSZ, low-redshift kSZ and radio sources from \citet{li22}. We show two different versions of the biases, one where we have assumed point sources with flux $>3\,\unit{mJy}$ at $145\,\unit{GHz}$ can be removed from the data, and a more futuristic case, using dashed lines, where sources sources with flux $>0.5\,\unit{mJy}$ at $145\,\unit{GHz}$ can be removed.  For comparison, the baseline 1$\sigma$ sensitivity to sources at this frequency for SO is 0.67 mJy \citep{abitol25}. 

The blue lines, for which $K$ is estimated from a minimum variance ILC, show that without further foreground mitigation, the bias due to foregrounds is one or two orders of magnitude above the signal. This is unsurprising - we know the tSZ and CIB fields are non-Gaussian and 
trace much of the same large-scale structure that generates CMB lensing. 

We show also versions which estimate $K$ from a harmonic ILC in which model spectra of the tSZ (orange lines), CIB (green lines) and of both the tSZ and CIB (purple lines) have been deprojected. Without the optimistic source subtraction, tSZ deprojection alone worsens the bias, likely because the linear combination of channels that nulls the tSZ increases the amplitude of the CIB and/or radio sources.
For the brown lines labelled ``one leg (tSZ + CIB)-deproj.", we perform the deprojection on only one of the temperature maps entering the $K$ estimator, which, in the case of perfect deprojection of foregrounds, would remove the same biases as performing the deprojection on both legs, but with a lower noise penalty (see \citealt{raghunathan23} and \citealt{maccrann24} for related approaches). We see that for the futuristic source threshold case (dashed lines), likely achievable for a lower noise experiment such as CMB-HD, biases due to extragalactic foregrounds can be reduced to order unity (indicated by the grey shaded region). 

Like the low-$z$ kSZ contamination discussed in \sect{sec:lowzsignal}, we note that the contamination could be further reduced by removing low redshift lensing from the $\phi$ estimate, by using, e.g., galaxy survey and CIB maps. As discussed in \sect{sec:lowzsignal}, a reduction in the contamination by a factor on the order of 5 should be possible using LSST galaxies. 

We note that as well imparting a bias, extragalactic foregrounds increase the variance in the observed temperature map, thus increasing  $\tilde{C}_l^{TT}$. In our baseline forecast, we have to some extent accounted for this by assuming a noise level  larger than that coming purely from the CMB, kSZ and instrumental noise components, appropriate for e.g. the case where these foregrounds have been deprojected via a constrained ILC (5\unit{\mu K arcmin} for the SO-like case, 0.5\unit{\mu K arcmin} for the CMB-HD-like case). Whether these experiments can achieve the statistical uncertainties assumed here of course depends  on whether they are able to control foregrounds to this level, which remains to be seen. 

It is likely that the rather simple foreground mitigation presented above could be improved in various ways. Additional frequency channels could be used e.g. from CCAT \citep{ccat-prime}, which will provide high-resolution, higher-frequency data that will be especially informative about the CIB. Given the uncertainty in the CIB frequency spectrum, extending the de-projection method via the moments-method \citep{chluba17,rotti21} may also be beneficial.
\citet{kusiak21,kusiak23} also present a range of more targeted foreground cleaning approaches, in particular for removing CIB contamination for kSZ$^2$ cross-correlated with galaxies, including a method called ``$\alpha$-cleaning", where an appropriately scaled high frequency map is subtracted from the temperature data such that the cross-correlation with the galaxy sample is nulled, as expected for a perfect blackbody map. Presumably a similar method could be applied to the cross-correlation with $\phi$.

\subsection{Lensing bias}\label{sec:lensingbias}

As discussed in the projected-fields literature (e.g. \citealt{hill16,bolliet22,kusiak21}), as well as in \citet{maccrann24} for the reionisation component, the $\hat{K}$ estimator is also sensitive to the mode-coupling induced by lensing.  Hence, our estimate $\hat{K}$ will have a bias
\begin{equation}
    \hat{K}_{\vecL} = K_{\vecL} + R_{L}^{K\phi}\phi_{\vecL}
\end{equation}
where $R^{K\phi}_{\vecL}$ quantifies the response to $\phi$ of the $K$ estimator. The cross-correlation $C_L^{K\phi}$ will then pick up a bias $R_{L}^{K\phi}C_L^{\phi\phi}$, which we refer to as the lensing bias.

\begin{figure}
    \centering
    \includegraphics[width=0.95\columnwidth]{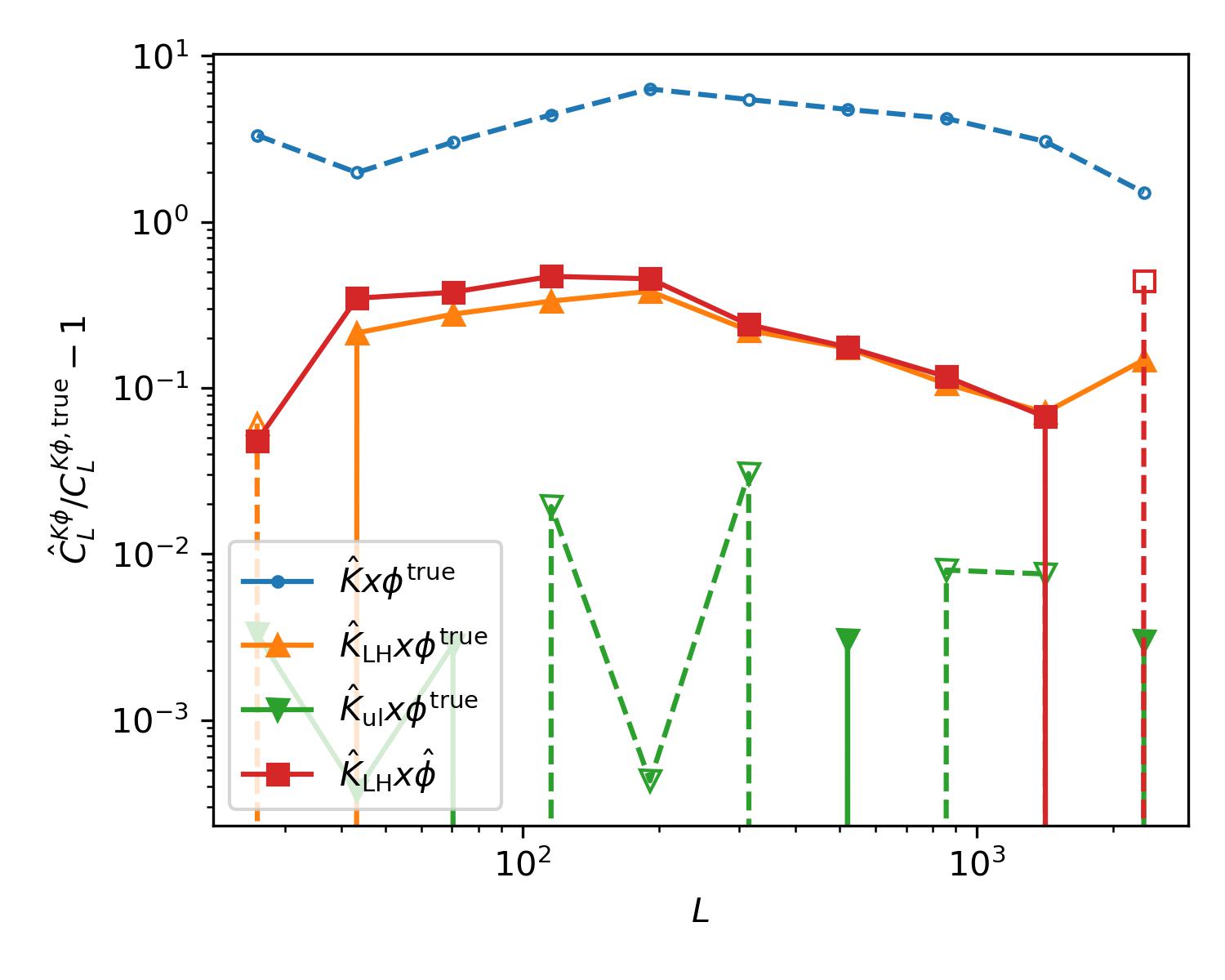}
    \caption{The fractional bias to $C_L^{K\phi}$ due to lensing contamination of the $\hat{K}$ estimator. Dashed lines with open symbols indicate a negative bias. The circles/blue line uses the normal $\hat{K}$ estimator, and picks up an order 1 fractional bias. For the green line/downward triangles, $K$ is estimated from a CMB temperature that is unlensed, hence this lensing bias vanishes. For the orange line/upwards triangles, the `lensing-hardened' $\hat{K}$ estimator is used, reducing the lensing bias to 10\%. When the estimtated $K$ is cross-correlated with the reconstructed $\phi$ (rather than the true $\phi$) there is only a small increase in the bias (red line/squares).}
    \label{fig:lensingbias1}
\end{figure}

If one assumes a cosmological model, and that $R^{K\phi}_{\vecL}$ is known (we will discuss this below), this bias can be modeled. However, given the size of the lensing bias, uncertainty in the $C_L^{\phi\phi}$ model may be non-negligible, especially if there is significant cosmic variance uncertainty on the realization of $C_L^{\phi\phi}$ in the data e.g. if as in \citet{raghunathan24}, one works with a small, very deep survey region in order minimise foreground contamination.
Following the development of bias-hardening for CMB lensing \citep{namikawa13,sailer20}, \citet{maccrann24} proposed bias-hardening against lensing contamination when estimating $K$. By noting that the $\phi$ estimator is also biased by the presence of $K$, one can write down bias-hardened $K$ and $\phi$ estimators, $K_{BH}$ and $\phi_{BH}$ by inverting 
\begin{equation}
\begin{bmatrix} \hat{K} \\ \hat{\phi} \end{bmatrix} = \begin{bmatrix} 1 & R^{K\phi} \\ R^{\phi K} & 1 \end{bmatrix} \begin{bmatrix} K \\ \phi \end{bmatrix}
\end{equation}
and calculating the response functions $R^{xy}$ following \citet{namikawa13,sailer20}. In \fig{fig:lensingbias1}, we show the fractional bias, in blue, due to lensing contamination of $\hat{K}$, again for the fiducial AMBER simulation. Note that we cross-correlate the estimated $K$ with the true (input) $\phi$; we discuss further biases which arise when using estimated $\phi$ below. For the green line, we replace the lensed CMB with unlensed CMB when estimating $K$, so the lensing bias vanishes. For the orange line, we use the lensing-hardened $K$ estimator, which reduces the fractional bias to $10\%$.

Some bias remains, which could arise from various approximations implicit in the lensing-hardening. Firstly, in calculating $R^{\phi K}$, the response of the $\phi$ estimator to $K$, one must assume an analytic form of the mode-coupling induced by $K$. We assume this mode-coupling has the form $\left<T_\vecl T_{\vecL-\vecl}\right> \propto \sqrt{C_l^{kSZ}}$ 
appropriate for Poisson-distributed blobs, when in reality the mode-coupling may be more complex.

Secondly, the lensing-hardened estimator only accounts for the linear response of the $K$-estimator to $\phi$. Especially given the high CMB $l$s used to estimate $K$, there may be some beyond-linear response present also. 

The picture will be complicated further since we are using a quadratic estimator to estimate the CMB lensing potential, where higher order biases will also be present, which may correlate with the higher-order biases in the lensing-hardened estimator. However, it appears this last effect is not very significant - the red line in \fig{fig:lensingbias1} correlates the lensing-hardened $K$ estimate, $\hat{K}_{\mathrm{LH}}$, with the $\hat{\phi}$ also estimated from the simulated CMB which results in only a small increase in the fractional bias. 

\section{Discussion}\label{sec.conclusion}

The epoch of reionisation, a key epoch in the Universe's history, is very difficult to observe directly, for example via 21cm emission. We can use the cosmic microwave background as a ``backlight" with which to view the EoR, via the kSZ effect - Doppler boosting of CMB photons by ionised electrons moving in the line-of-sight direction. 

While the signatures of reionisation in the power spectrum and trispectrum of the kSZ temperature perturbation are well-studied, we proposed a new statistic here: the cross-correlation of $K$ (the squared, small-scale CMB temperature) with the lensing potential $\phi$. This complements similar statistics aiming to probe reionisation such as $C_L^{K\tau}$ \citep{kramer25} and the cross-correlation of $K$ with galaxies \citep{La_Plante_2022}. It is  the high-redshift equivalent to what is known as the ``projected-field'' kSZ statistic, which has been extensively studied for the low-redshift kSZ \citep{dore2004,hill16,ferraro16,kusiak21,bolliet22,kusiak23}.
We have argued that $C_L^{K\phi}$ probes the bispectrum, $\left<\delta_e \delta_e \delta_m\right>$, during reionisation, and  is sensitive to the midpoint and duration of reionization. 

$C_L^{K\phi}$ can be expressed as a CMB trispectrum, since both $K$ and $\phi$ can be estimated using quadratic estimators on the observed CMB. We discussed various challenges involved with estimating this trispectrum, including the ``$N^0$ bias'', which can be removed e.g. by using only polarisation data to estimate $\phi$. The biases due to the low-redshift kSZ and  other extragalactic foregrounds are more challenging but all would be significantly mitigated by ``cleaning'' of the low-redshift lensing using galaxy catalogs or maps of the CIB. Our initial investigations here suggest these biases could be reduced to around the level of the expected signal - but we are optimistic that further improvements could be possible with foreground-cleaning methods that are more tailored to this statistic (e.g. as explored in \citealt{kusiak21,kusiak23} for the low-redshift kSZ signal).

While the $C_L^{K\phi}$ signal appears to be only marginally detectable with the Simons Observatory, a futuristic experiment like CMB-HD could make a $S/N\sim50$ detection, and thus clearly discriminate between different models of reionisation. It joins a host of other exciting potential CMB secondaries measurements that would be sensitive to reionisation physics with a futuristic low-noise, high-resolution CMB experiment like CMB-HD. As well as the kSZ and $\phi$ during the  EoR, cross-correlations with patchy screening \citep{feng19,bianchini23,kramer25} may further elucidate the picture. A particularly promising direction for future work is the combination of the kSZ power spectrum, the kSZ trispectrum, $C_L^{KK}$, and $C_L^{K\phi}$ which is likely to be provide constraints on the EoR via breaking degeneracies between reionisation physics parameters \citep{alvarez20}, and their different sensitivites to extragalactic foregrounds. 

\section*{Acknowledgements}

NM is supported by a Royal Society University Research Fellowship. AvE thanks the Kavli Institute for Cosmology Cambridge for their hospitality during his stay as a Kavli Medium-Term Visitor,  during which part of this work was performed.  AvE and DK were supported by NASA grant 80NSSC24K0665.  AvE and DK were additionally supported by NASA grants 80NSSC23K0747 and 80NSSC23K0464, and NSF AAG grant 588167. FMcC acknowledges support from the European Research Council
(ERC) under the European Union’s Horizon 2020 research and innovation programme (Grant agreement
No. 851274). CC acknowledges support from the Beus Center for Cosmic Foundations at Arizona State University
\section*{Data Availability}

This paper uses already publicly available data. 



\bibliographystyle{mnras}
\bibliography{refs} 








\bsp	
\label{lastpage}
\end{document}